\documentstyle[12pt,aasms4]{article}

\lefthead{Sako et al.}
\righthead{Very High-Energy Gamma-rays from PSR B1509$-$58}

\begin{document}

\title{Very High-Energy Gamma-Ray Observations of PSR B1509$-$58
with the CANGAROO 3.8m Telescope}

\author{T. Sako             \altaffilmark{1,14},
        Y. Matsubara        \altaffilmark{1},
        Y. Muraki           \altaffilmark{1},
        P. V. Ramanamurthy  \altaffilmark{1},
        S. A. Dazeley       \altaffilmark{2},\\
        P. G. Edwards       \altaffilmark{3},
        S. Gunji            \altaffilmark{4},
        T. Hara             \altaffilmark{5},
        S. Hara             \altaffilmark{6},
        J. Holder           \altaffilmark{7, 15},
        S. Kamei            \altaffilmark{6},
        A. Kawachi          \altaffilmark{7},\\
        T. Kifune           \altaffilmark{7},
        R. Kita             \altaffilmark{8},
        A. Masaike          \altaffilmark{9},
        Y. Mizumoto         \altaffilmark{10},
        M. Mori             \altaffilmark{7},
        M. Moriya           \altaffilmark{6},\\
        H. Muraishi         \altaffilmark{8},
        T. Naito            \altaffilmark{10},
        K. Nishijima        \altaffilmark{11},
        S. Ogio             \altaffilmark{6},
        J. R. Patterson     \altaffilmark{2},
        G. P. Rowell        \altaffilmark{7},\\
        K. Sakurazawa       \altaffilmark{6},
        Y. Sato             \altaffilmark{7},
        R. Susukita         \altaffilmark{12},
        R. Suzuki           \altaffilmark{6},
        T. Tamura           \altaffilmark{13},\\
        T. Tanimori         \altaffilmark{6},
        G. J. Thornton      \altaffilmark{2},
        S. Yanagita         \altaffilmark{8},
        T. Yoshida          \altaffilmark{8} 
    and T. Yoshikoshi       \altaffilmark{7}}

\altaffiltext{1}{Solar-Terrestrial Environment Laboratory, Nagoya University, Nagoya 464-8601, Japan}
\altaffiltext{2}{Department of Physics and Mathematical Physics, University of Adelaide, South Australia 5005, Australia}
\altaffiltext{3}{Institute of Space and Astronautical Science, Sagamihara 229-8510, Japan}
\altaffiltext{4}{Department of Physics, Yamagata University, Yamagata, Yamagata 990-8560, Japan}
\altaffiltext{5}{Faculty of Management Information, Yamanashi Gakuin University, Kofu 400-8575, Japan}
\altaffiltext{6}{Department of Physics, Tokyo Institute of Technology, Tokyo 152-8551, Japan}
\altaffiltext{7}{Institute for Cosmic Ray Research, University of Tokyo, Tokyo 188-8502, Japan}
\altaffiltext{8}{Faculty of Science, Ibaraki University, Mito 310-8521, Japan}
\altaffiltext{9}{Department of Physics, Kyoto University, Kyoto 606-8502, Japan}
\altaffiltext{10}{National Astronomical Observatory of Japan, Tokyo 181-8588, Japan}
\altaffiltext{11}{Department of Physics, Tokai University, Hiratsuka 259-1292, Japan}
\altaffiltext{12}{Institute of Physical and Chemical Research, Wako 351-0198, Japan}
\altaffiltext{13}{Faculty of Engineering, Kanagawa University, Yokohama 221-8686, Japan}
\altaffiltext{14}{present address; LPNHE, Ecole Polytechnique, Palaiseau CEDEX 91128, France; sako@poly.in2p3.fr}
\altaffiltext{15}{present address; LAL, Universite de Paris-Sud, BP 34, ORSAY CEDEX 91898, France}

\begin{abstract}
The gamma-ray pulsar PSR B1509$-$58 and its surrounding nebulae have been
observed with the CANGAROO 3.8m imaging atmospheric \v{C}erenkov telescope.
The observations were performed from 1996 to 1998 in Woomera, 
South Australia, under different instrumental conditions with
estimated threshold energies of 4.5 TeV (1996), 1.9 TeV (1997) 
and 2.5 TeV (1998) at zenith angles of $\sim$ 30$^{\circ}$.
Although no strong evidence of the gamma-ray emission was found,
the lowest energy threshold data of 1997 showed a marginal excess of 
gamma-ray--like events at the 4.1 $\sigma$ significance level.
The corresponding gamma-ray flux is calculated to be
$\rm (2.9\,\pm\,0.7) \times 10^{-12}\,cm^{-2}s^{-1}$ 
above 1.9 TeV.
The observations of 1996 and 1998 yielded only upper limits (99.5\% 
confidence level) of
$\rm 1.9 \times 10^{-12}\,cm^{-2}s^{-1}$ above 4.5 TeV
and
$\rm 2.0 \times 10^{-12}\,cm^{-2}s^{-1}$ above 2.5 TeV, respectively.
Assuming that the 1997 excess is due to Very High-Energy (VHE)
gamma-ray emission from the pulsar nebula,
our result, when combined with the X-ray observations, leads to a value of the
magnetic field strength $\simeq$ 5 $\mu$G.
This is consistent with the equipartition value previously estimated in the
X-ray nebula surrounding the pulsar.
No significant periodicity at the 150\,ms pulsar period has been 
found in any of the three years' data.
The flux upper limits set from our observations are one order of magnitude
below previously reported detections of pulsed TeV emission.
\end{abstract}


\section{Introduction} \label{INTRO}
Pulsar nebulae have been suggested as a possible acceleration site of 
high-energy particles in the galaxy (\cite{har90}).
The first order Fermi acceleration mechanism is expected to occur in a
shock between the pulsar wind and supernova ejecta, or interstellar matter.
Evidence of such energetic phenomena has been obtained through observation of
synchrotron emission by accelerated electrons and positrons at radio to 
gamma-ray ($\leq$ 10 GeV) energies.
However, more direct evidence has become obtainable through Very 
High-Energy (VHE) gamma-ray ($\geq$ 300 GeV) observations over the last decade
using the Imaging Atmospheric \v{C}erenkov Technique (IACT).

VHE gamma-ray emissions from the directions of three energetic pulsars,
the Crab (\cite{wee89}; \cite{vac91}; \cite{tan94});
the Vela pulsar (\cite{yos97}) and PSR B1706$-$44 (\cite{kif95}, \cite{cha97}),
have been detected by ground-based telescopes using the IACT.
Although all three pulsars show pulsed emission in the {\it EGRET} energy
range (100MeV--10GeV), none of the VHE gamma-ray detections have shown any
periodicity at the radio pulsar period.
This steady VHE gamma-ray emission is usually explained to be a result of the 
inverse Compton scattering in the pulsar nebula, and not from the pulsar
magnetosphere.
While the mechanism of the emission from the Crab nebula is well studied
(see, for example, \cite{jag96}), information on other pulsars is still
sparse.
In order to study pulsars and their surrounding environment
as possible acceleration sites of the cosmic rays, more examples in the VHE
gamma-ray range are required.

PSR B1509$-$58 was discovered as an X-ray pulsar by Seward and Harnden (1982)
using the {\it Einstein X-ray Observatory}.
It is near the center of the supernova remnant MSH15$-$52 (G320.4$-$1.2).
Soon after this discovery, pulsed radio emission was found by 
Manchester, Tuohy and D'Amico (1982).
The pulsar has a period of 150 msec and a period derivative of 
1.5 $\times$10$^{-12}$ ss$^{-1}$, the largest known today. 
The characteristic age of the pulsar is 
estimated to be $\sim$1700 years (\cite{man98}), which makes it the second 
youngest pulsar after the Crab.
\altaffilmark{1}
\altaffiltext{1}
{ Torii et al.\ (1997) have reported the discovery of a pulsar 1600 years
old.
This age is somewhat speculative however as the period derivative of the 
pulsar has not yet been measured and
an association with a historical supernova was assumed to estimate
the pulsar age.}
From the period and the large period derivative, a very strong
surface magnetic field of 1.5 $\times$ 10$^{13}$ G and a large spin down
energy loss rate of 1.8 $\times$ 10 $^{37}$ ergs s$^{-1}$ are implied.
While the distance to the pulsar is relatively large (4.4 kpc, \cite{tay95}),
the expected energy flux received at the Earth is the fifth largest among the
known pulsars.

A compact ($\rm\sim\,10^{\prime}\,\times\,6^{\prime}$) synchrotron X-ray nebula
has been found to exist around PSR B1509$-$58 (\cite{sew84}).
The synchrotron emission suggests the existence of non-thermal electrons
(positrons) in the nebula, which will also emit VHE gamma-rays 
via inverse Compton scattering. 
A detectable VHE gamma-ray flux from this synchrotron nebula was predicted
by du Plessis et al.\ (1995) as a function of the magnetic field strength in
the nebula.
The expected gamma-ray flux above 1 TeV of 10$^{-11}$ to 
10$^{-12}\,cm^{-2}s^{-1}$ for nebula magnetic fields 4 to 10 $\mu$G is within
the sensitivity of the CANGAROO 3.8m telescope.
Thus, VHE observations should give a good measurement of the magnetic field
strength of this nebula.
Du Plessis et al.\ (1995) also predicted a very hard differential spectral 
index of $\sim$1.8 based on the X-ray observations.
This prediction provides us with an extreme example of the utility of
multiwavelength studies of synchrotron---inverse- Compton emitting objects.
Besides the compact nebula, recent X-ray satellite observations suggest
various non-thermal phenomena in this remnant.
{\it ROSAT} observations indicate a non-thermal X-ray component from the
central diffuse nebula (CDN) extending to a diameter of 50$^{\prime}$ 
($\sim$ 60pc) centered on the pulsar (\cite{tru96}).
{\it ASCA} observations revealed a non-thermal jet structure between the 
pulsar and the center of a thermal nebula about 10$^{\prime}$ north from the
pulsar (\cite{tam96}).
In order to explain the effective thermalization process of the thermal nebula,
Tamura et al.\ (1996) indicate the existence of accelerated ions as well as
electrons in the jet.
Furthermore, Gaensler et al.\ (1998) found synchrotron emission from compact 
knots in this thermal nebula from 20cm imaging observations with the
Australia Telescope Compact Array.

The surface magnetic field strength of the pulsar PSR B1509$-$58 is estimated 
to be one of the largest among known pulsars.
Due to the photon splitting process caused by this strong surface magnetic 
field, a cut-off in the pulsed emission around MeV energies is predicted
by Harding, Baring and Gonthier (1997).
In fact, Kuiper et al.\ (1999) have suggested that a cut-off around 10 MeV
exists in the {\it COMPTEL} data.
{\it EGRET} observations have resulted in only an upper limit for the pulsed 
emission from PSR B1509$-$58 (\cite{tho94}).
In contrast, Nel et al.\ (1992) have reported the detection of transient 
pulsed VHE gamma-rays from the observations between 1985 and 1988
based on ground--based (non-imaging) \v{C}erenkov telescope observations.
However they could not detect any significant pulsed emission in the
successive years.
They tried to explain their observations with the framework of the outer gap
model (\cite{chr86}).
(Bowden et al.\ (1993) reported a upper limit of the pulsed emission above
0.35 TeV from their observations in 1987 and 1989.
Combining with the detection by Nel et al.\ (1992) in 1987 above 1.5 TeV,
power law index of the integral energy spectrum is limited to be harder than
$\sim$ 1.)
Interestingly, Kuiper et al.\ (1999) also indicate a marginal detection of the
pulsed emission above 10 MeV, where the origin may
differ from that at lower energies.
Consequently, we have examined our data for the presence of periodicity as 
well.
Our observations are the first results on this pulsar with using the
IACT, which is one order of magnitude more sensitive than 
non-imaging observations.

For the reasons given above, we believed that PSR B1509$-$58 would be an
interesting object to study above 1 TeV energies with the CANGAROO 3.8m
IACT telescope in both the steady nebula emission and the pulsed emission.
Details of those observations are given in Section-\ref{OBS}.
The methods of the analysis and results are shown in Section-\ref{ANA}. 
In Section-\ref{DIS}, we summarize our results and discuss their
implications.

\section{Observations} \label{OBS}
The CANGAROO (Collaboration between Australia and Nippon (Japan) for a
GAmma-Ray Observatory in the Outback) 3.8m telescope is located at Woomera,
South Australia
(136$^{\circ}$47$^{\prime}$E, 31$^{\circ}$6$^{\prime}$S and 160m a.s.l.).
\v{C}erenkov photons emitted from extensive air showers originated by
primary gamma-rays and cosmic rays are collected with a parabolic mirror of
3.8m diameter and detected with an imaging camera at the focal plane.
The camera consists of 256 photomultiplier tubes (PMTs) of 10mm$\times$10mm
size (Hamamatsu R2248).
The PMTs are located in a 16$\times$16 square grid and the field of view
amounts to 3$^{\circ} \times$ 3$^{\circ}$.
When signals from more than 5 tubes exceed 3 photoelectrons each within a 
gate, a trigger is generated.
The amplitude and relative time of each PMT signal, the event time,
and the counting rate of each tube are recorded for each event.
The absolute time can be obtained with a precision of 200 nsec using a GPS
clock.
In addition to the GPS clock, the time of a crystal clock with a precision of
100$\mu$sec is also recorded.
The GPS clock was not available in the 1997 observations due to the
installation work of our new data acquisition system.
However, because the time indicated by the crystal clock shows a stable drift
from that of the GPS clock, we can obtain accurate {\it relative} arrival 
times for events even without the GPS clock.
The crystal clock is reset every observation (new moon) period.
Therefore, a periodicity analysis based on this clock is valid on a 
month by month basis.
GPS timing was restored in July 1997.
Details of the camera and the telescope are described in Hara et al.
(1993).

The telescope was pointed in the direction of the pulsar PSR B1509$-$58
(right ascension 15$^{h}$13$^{m}$55$^{s}$.62 and declination
--59$^{\circ}$ 08$^{\prime}$ 08$^{\prime\prime}$.9 (J2000), \cite{tay95})
in May and June in 1996, from March to May in 1997 and from March to May in
1998.
The pulsar (ON source) and an offset region (OFF source), having the same 
declination as the pulsar but different right ascension, were observed for
equal amounts of time each night under moonless and usually clear sky 
conditions.
Typically, the ON source region is observed only once in a night around 
transit for a few hours.
Two OFF source runs are carried out before and after the ON source run.
The former one covers the first half of the ON source track and
the latter covers the second half.
In the off-line analysis, those data obtained when a small patch of cloud was
obscuring the source are omitted.
At the same time, the corresponding ON (or OFF) source data 
were also rejected from the analysis.
In addition to the weather selection, the data taken when the electronics
noise produced an anomalously large trigger rate were not used in the analysis.
This happened in the 1996 observations.
In the 1998 data, there are many nights which have a large difference of the 
event rate between the ON and OFF source regions, which is thought to be due to the 
presence of thin dew on the reflecting mirror.
Data taken under these conditions were also omitted. 
The durations of selected observations after these procedures are
$\rm 26^{h}30^{m}$ , $\rm 32^{h}08^{m}$ and $\rm 21^{h}14^{m}$
for the 1996, 1997 and 1998 (both ON and OFF) data, respectively.
These data are used for the analysis in this paper.

Observations were carried out under different instrumental conditions in each
year.
During the 1996 observations, the reflectivity of the mirror was estimated to
be $\simeq$ 45\%.
We recoated the mirror in October 1996 by vacuum evaporation of aluminium at
the Anglo Australian Observatory.
As a result, the reflectivity of the mirror increased to about 90\%.
As the reflectivity was improved, the threshold energy of our telescope 
was lowered.
For the 1997 observations, the threshold energy,
defined here as the energy at which a
differential photon flux with an assumed differential spectral index of 2.5
is maximized 
in the Monte Carlo calculations, was estimated to be 1.9 TeV,
compared to 4.5 TeV before the recoating.
By the 1998 observations, the reflectivity had decreased to $\simeq$ 70\%, 
corresponding to a threshold energy of 2.5 TeV.
In these estimations, the selection effect of the analysis described
in the next section is also taken into account.
In the Monte Carlo calculation, we assumed that the observations were made at 
a zenith angle of 30$^{\circ}$, which was close to the average value for our 
observations on PSR B1509$-$58.

The observation times and threshold energies are summarized in 
Table~\ref{obssum}, and as well, the analysis results are shown.

\placetable{obssum}

\section{Analysis and Results} \label{ANA}
\subsection{Analysis method}
At the beginning of each run, the ADC pedestal and gain for each PMT were
measured.
To calibrate the gain, a blue LED located at the center of the mirror is used
to illuminate the PMTs uniformly.
The pedestal value is subtracted from the ADC value and any
variations in the PMT gains were normalized using the LED calibration data.
PMTs whose TDC value corresponded to a pulse arrival time within 
$\pm$ 30 nsec of the shower plane were regarded as \lq hit \rq ~tubes and used 
to calculate image parameters.
After omitting some hit tubes which were isolated or which had ADC values
less than one standard deviation above the pedestal value, the conventional
image parameters (\cite{hil85}) were calculated.
(In the 1996 data, a fifth of the PMTs at the bottom in the camera were
omitted from analysis to avoid the effect of electronics noise.
This makes the threshold energy higher and the effective area smaller.
This effect is included in calculating the threshold energy and the flux 
upper limit.)

The parameter ranges determined from Monte Carlo simulations to optimize
the gamma-ray signals are $\colon$
$0^{\circ}.60<$ {\it distance} $\leq1^{\circ}.30$,
$0^{\circ}.04<$ {\it width} $\leq 0^{\circ}.09$,
$0^{\circ}.10<$ {\it length} $\leq0^{\circ}.40$, 
$0.35<$ {\it concentration} $\leq0.70$ 
and $\alpha\,\leq10^{\circ}$.
These ranges are slightly narrower than those used in case of the Vela
analysis (\cite{yos97}).
The upper limit of $\alpha$, $10^{\circ}$, is adopted assuming the source is
a point-like.
Two orientation parameters, $\alpha$ and {\it distance}, are defined with
respect to the assumed source position in the field of view.
In this paper, this is fixed at the pulsar position except in the spatial 
analysis discussed in Section-\ref{MAP}.
To avoid the effect of incomplete images near the edge of the camera,
images with centroids located at greater than $1^{\circ}.05$ from the center
of the camera were also rejected.
We also required that the number of hit tubes ($N_{hit}$) must be $\geq$
5 and the total number of photo-electrons contained in an image ($N_{p.e.}$)
must be $\geq 40$ to be able to obtain good image parameters and select
only air shower induced events.
The upper limit of $N_{p.e.}$ is large enough to accept all real
events with large numbers of photo-electrons.
In Table~\ref{obssum}, the numbers of events in the raw data and selected
are presented.
We can find a large difference between ON and OFF in the raw data.
The main reasons of the difference in number are the electronics noise 
in the 1996 data and the existence of the optically bright stars 
(M$_{V}$ = 4.1 and 4.5) in the field of view in the 1997 data,
where the reflectivity of the mirror was the largest.
However, the numbers match well after the selection of air shower events. 
For all the three years' data analyses we applied the same criteria as
described above.

\subsection{Results of the image analysis}
The distributions of the orientation angle ($\alpha$) after all other cuts
were applied are shown in Figure~\ref{Fig1}.

\placefigure{Fig1}

Although there was no statistically significant excess of the ON source counts
over the OFF source seen in the 1996 data, the 1997 data shows an excess at 
$\alpha\leq10^{\circ}$ with a statistical significance of 4.1$\sigma$.
This excess may indicate the presence of a VHE gamma-ray signal from the 
source.
The additional use of the {\it asymmetry} parameter showed an
excess in the positive (gamma-ray--like) domain, though not at a level
which would have increased the overall significance of the excess.
More careful study would be necessary in use of this third-moment
parameter for the source near the Galactic Center, where the night-sky
background level is high.
In the 1998 data, we find a small excess in the ON source counts, however,
the statistical significance is only 1.4$\sigma$ at $\alpha\leq10^{\circ}$.
Hereafter, we regard the 1996 and 1998 results as non-detections of the VHE
gamma-ray signal and treat the 1997 result as a marginal detection.
The corresponding upper limits and flux are calculated as,

$\rm F_{99.5\%}(E\,\geq\,4.5\,TeV)\,\leq\,1.9 \times 10^{-12}~cm^{-2}~s^{-1}$

$\rm F(E\,\geq\,1.9\,TeV)=(2.9\,\pm\,0.7) \times 10^{-12}~cm^{-2}~s^{-1}$

$\rm F_{99.5\%}(E\,\geq\,2.5\,TeV)\,\leq\,2.0 \times 10^{-12}~cm^{-2}~s^{-1}$

\noindent
for 1996, 1997 and 1998 results, respectively.
Here, a differential energy spectral index of 2.5 is assumed.
The upper limits and the errors in the flux are estimated based on the numbers
of the observed counts.
We note that in our calculation of the upper limits the difference of
the counts between ON and OFF are also taken into account following the
formula introduced by Helene (1983).
So the 1998 flux upper limit becomes higher than that from completely
null result.
If we change the assumption of the differential energy spectral index over
the range 
$2.5~\pm~1.0$, the corresponding threshold energies are estimated to change
by $\sim\mp$30\%.
Instrumental uncertainties also affect the estimation of the threshold
energies.
We estimate the systematic error in determining the absolute threshold 
energies to be about 40$\sim$50\%.
However, because almost all of the systematic errors behave in the same way 
for the three years' observations, the uncertainty of the relative threshold
energy is smaller than this value.

\subsection{Consistency and Stability}
The positive indication is obtained only from the lowest threshold energy 
observation.
But the derived flux and two flux upper limits require neither variability 
of the source nor a very soft spectral index, that is, the results from the
three years are consistent with each other assuming stable emission with a 
Crab-like spectral index ($\sim 2.5$) or the harder index (1.8) expected
by du Plessis et al.\ (1995). 
We also divided the 1997 data into separate new moon periods to check on
consistency.
The results are shown in Table~\ref{obssum}.
Each month's result has a marginal positive effect on the final result.
The excess counting rate is stable during the three observation seasons
within the statistical errors.

\subsection{Spatial analysis} \label{MAP}
PSR B1509$-$58 and its surrounding environment are complex and there are
indications from X-ray data that non-thermal phenomena possibly occur 
over an extended area of this remnant.
So it is possible that the gamma-ray--like signal in the 1997 data is not from
a point source at
the pulsar position but from some other region near the pulsar.
Therefore we have carried out a source search in the $2^{\circ}\times2^{\circ}$
field of view centered on the pulsar position.
To do this, we shifted the position of the assumed source over a grid of points
around the pulsar and repeated the analysis at each point to obtain the excess
counts in the $\alpha$ distribution.
The resultant map of the significance is shown in Figure~\ref{Fig4}.

\placefigure{Fig4}

The peak of the excess is found at $0^{\circ}.1$ south-west from the pulsar.
But when we consider the degrees of freedom of the search, the significance
at this maximum should be reduced. 
And also, from a Monte Carlo calculation, where the observed counts of signal
and background are taken into account, we estimated that the precision to 
determine the source position is $0^{\circ}.10$ at the 1$\sigma$ level.
We conclude, therefore, that the position of the excess is consistent with 
the pulsar position within the statistics of our observations.

\subsection{Periodicity analysis}
The recorded arrival times of the gamma-ray--like signals ($\alpha \le
10^{\circ}$ after all the image cuts) were converted to the Solar System
Barycenter arrival times using the solar system ephemeris based on
epoch 2000 (DE200) (Standish, 1982).
We then carried out a phase analysis with the phase parameters summarized 
in Table~\ref{tbl-eph} (\cite{man98}).
Because Nel et al.\ (1990) pointed out a possibility of a light curve with 
triple peaks in the TeV energy range, we applied the H-test (\cite{jag89})
to obtain the statistical significance.
The virtue of the H-test lies in the fact that it 
requires no assumptions about bin size and bin location and is 
also independent of the shape of the
light curve.
The results are summarized in Table~\ref{tbl-1}.

\placetable{tbl-1}

The results of 1997 are divided into separate observational periods (months), 
because GPS timing information was not available in 1997 as mentioned in 
Section-\ref{OBS}. 
The relative arrival time of the events is calculated for the 1997 data from 
the time of the crystal clock, having a constant drift rate 
relative to the GPS clock.
The H-statistics and the corresponding probabilities against a uniform
distribution are shown in Table~\ref{tbl-1}.
No evidence for the 150\,ms periodicity is found in any of the observation 
seasons. 
To calculate the flux upper limit for the pulsed emission, we used the formula
given by de Jager (1994).
This formula combines the observed counts (N) and pulsed fraction (p) through
a parameter, $\chi$, as, $\chi = p \sqrt{N}$.
When the H-statistic is considered as a non-detection of periodicity, $\chi$
giving 3 $\sigma$ upper limit of p is expressed as,

\[ \chi_{3\sigma} = ( 1.5 + 10.7 \delta ) ( 0.174 H ) ^{0.17+0.14\delta} 
   exp \left[ ( 0.08 + 0.15\delta ) \left\{ log_{10}(0.174H) \right\}^{2} 
   \right]
\]

Here, H is the value of the H-test as shown in Table~\ref{tbl-1}.
(For H$<$0.3 we should take H=0.3 in calculating $\chi_{3\sigma}$.)
$\delta$ is the duty cycle of the pulse profile.
In case of PSR B1509$-$58, we assumed $\delta$ to be 0.3 using the X-ray
observation by Kawai et al.\ (1991).
The 3$\sigma$ upper limits for the pulsed VHE gamma-ray emission are also shown
in Table~\ref{tbl-1}.

\section{Discussion} \label{DIS}
Our observations can be summarized as follows $\colon$ 
(1) In the observations with the lowest detection threshold energy, 
a 4.1$\sigma$ excess of gamma-ray--like events is found.
Null results in the observations of the other years (when the detection
threshold energies were higher) are not in conflict with
this marginal positive result $\colon$
neither variability of the source nor an especially soft energy spectrum
needs to be invoked.
(2) From the result in the 1997 observations, there is no evidence
of a variability on a monthly time-scale during three observation seasons.
(3) In the 1997 data, the peak emission source position is shifted slightly 
to the  south-west direction from the pulsar position.
However, considering the statistical error including the real event numbers
observed, this is consistent with the pulsar position.
(4) The periodicity of the events modulated with the radio pulsar period 
is studied.
We found no evidence of the 150~ms pulsar periodicity using the H-test 
in any of the observations for three years.

The statistical significance of the 1997 excess, 4.1$\sigma$, is too small
to claim as the detection of a VHE gamma-ray source, however, 
it is sufficiently suggestive to allow discussion supposing 
the excess was due to a VHE gamma-ray signal.
With this scheme
the simplest and most straightforward explanation can be made assuming 
that the emission is found from the pulsar nebula surrounding the pulsar.
VHE gamma-ray emission from a pulsar nebula is usually considered as a result
of inverse Compton scattering by relativistic electrons.
From the emission processes of synchrotron and inverse Compton radiations, a 
simple equation, 
$\rm \frac{\dot{E}_{synch}}{\dot{E}_{iC}}=\frac{\epsilon_{B}}{\epsilon_{ph}}$,
can be obtained.
Here $\rm \dot{E}_{synch}$ and $\rm \dot{E}_{iC}$ are the luminosities through
synchrotron radiation (mainly resulting in quanta in the X-ray energy range)
and inverse Compton scattering (mainly producing VHE gamma-rays), respectively,
and $\rm\epsilon_{B}$ and $\rm\epsilon_{ph}$ are the energy densities of the 
magnetic field and the target photons for inverse Compton scattering at the
emission region.
Assuming isotropic emission of both X-rays and gamma-rays,
$\rm \frac{\dot{E}_{synch}}{\dot{E}_{iC}}$ can be equated to
$\rm \frac{F_{synch}}{F_{iC}}$.
Here $\rm F_{synch}=7.2\times10^{-11}~ergs~cm^{-2}\,s^{-1}~(0.1-2.4\,keV)$
as given by Trussoni et al.\ (1996) and 
$\rm F_{iC}=2.7\times10^{-11}~ergs~cm^{-2}\,s^{-1}$, 
obtained by integrating the 1997 flux
above 1.9\,TeV assuming a 
differential spectral index of 2.5.
(The corresponding luminosity at the pulsar, $L_{iC}$, is 
$\rm 6.2\times10^{34}~ergs~s^{-1}$ assuming the pulsar distance of 4.4 kpc.
That is 0.34\% of the pulsar rotating energy loss.)
If the 3\,K Microwave Background Radiation (MBR) is the only target of the 
inverse Compton radiation, {\it i.e.},
$\rm \epsilon_{ph}=\epsilon_{3K}=3.8\times10^{-13}~ergs~cm^{-3}$, 
one obtains
$\rm \epsilon_{B}= 1.0\times10^{-12}~ergs~cm^{-3}$.
This, then, leads to a value for the magnetic field strength
$\rm B\,\simeq\,5\,\mu G$.
Considering the large uncertainties in the arguments above, this value agrees
well with the previously estimated value of $\rm B\,\simeq\,7\,\mu G$,
from the equipartition of energy between the particles and the magnetic
field (\cite{sew84}).
According to the prediction of du Plessis et al.\ (1995), our result 
corresponds to a magnetic field strength of $\rm B~\simeq~5~\mu G$.
These three estimated values of the magnetic field agree very well with each
other.

An alternative source of the target photons is the IR source IRAS 15099$-$5856,
known to be positionally coincident with the pulsar (\cite{are91}).
Du Plessis et al.\ (1995) estimated that the contribution from the IR photons
to the VHE gamma-ray flux would be at the same level as that from the 3\,K MBR.
However the association between IRAS 15099$-$5856 and the pulsar is uncertain.
In case that the IRAS source found at 25$\mu$m supplies the target photon for 
the inverse Compton process, the resultant VHE gamma-ray spectrum is expected
to be softer than that made from the 3\,K MBR.
This is because the critical energy of the parent electrons in the 
Klein-Nishina cross section is $\sim~6~\times~10^{12}$ eV against 25$\mu$m
IR radiation while it is $\sim~10^{15}$ eV for the 3\,K MBR.
Therefore, the VHE gamma-ray spectrum should have a rapid softening over the 
TeV energy range.
To understand the association of this IRAS source, detailed spectral
measurements with future observations are required as well as the X-ray 
observations discussed below.

While our observations do not place any interesting limit on the spectral
index, the very hard spectrum predicted by du Plessis et al.\ (1995)
should be discussed.
Their prediction was based on the observational
results of the X-ray spectrum which showed a hardening of the index in the
energy range below a few keV (photon index 1.4$^{+0.4}_{-0.2}$ below 4\,keV  
while 2.15$\pm$0.02 between 2 keV and 60 keV).
However, recent X-ray observations do not confirm this hardening.
The photon indices obtained in the wide X-ray energy band are consistent with 
a value around 2.2 (\cite{tru96}; \cite{tam97}; \cite{mar97}) though
the error of the ROSAT result is large.
To discuss the synchrotron spectrum in detail, we need information from 
radio observations. 
But, even with the recent high resolution observations, a radio pulsar wind 
nebula has not been discovered (\cite{gae98}).
 
The upper limits set to the periodic signal in this paper are one order of
magnitude below the previously reported flux in the same energy band 
(\cite{nel92}).
Although Nel et al.\ reported upper limits from observations after 1988,
our results should provide a far stricter limit on models.
The VHE pulsed emission is in conflict with the observed cut-off around 10
MeV as predicted by the polar-cap model.
To explain the VHE pulsed emission, an additional hard component,
probably outer-gap emission, is required.
Future observations by GLAST may reveal the existence of this component and
studies of its flux and spectral variability may hint at 
large variability in the VHE range.
The flux of the transient VHE pulsed emission reported in 1985,
$\rm F(E\,\geq\,1.5\,TeV)=(3.9\,\pm\,0.9) \times 10^{-11}~cm^{-2}~s^{-1}$,
would make this source the brightest known VHE gamma-ray source in the 
southern hemisphere.
We could detect this kind of activity even with short duration monitoring.
Semi-simultaneous monitoring of this pulsar with the future large IACT 
arrays in the southern hemisphere (CANGAROO-III, HESS) and GLAST
would be of great interest if the pulsar were to display such an
active phase in the future.

Finally, it is notable that, unlike the other pulsar nebulae detected at
VHE energies, PSR B1509$-$58 is not firmly detected by 
{\it EGRET} onboard the {\it CGRO} satellite.
In contrast, this pulsar and its surroundings show a variety of the
non-thermal phenomena as introduced in Section-\ref{INTRO}.
A comparison of nonthermal X-ray emission with VHE gamma-ray emission is
becoming very useful in the search for VHE gamma-ray sources and study of 
their environment.
Combined with the recent studies of pulsar nebulae (\cite{kaw96}), 
the new generation of the Imaging Atmospheric \v{C}erenkov Telescopes
(e.g.\ \cite{mat97}) will result in an improved understanding of pulsar 
nebulae and particle acceleration.
The CANGAROO~II 7m telescope started observations at Woomera in mid-1999.
From new observations with a lower energy threshold, we will be able to 
measure the
gamma-ray spectrum precisely and obtain a better estimation of the 
physical parameters, especially the magnetic field strength, 
in pulsar nebulae. 

\acknowledgments
This work is supported by a Grant-in-Aid in Scientific Research from the Japan
Ministry of Education, Science, Sports and Culture, and also by the Australian
Research Council.
We would like to thank to Dr. R. N. Manchester who provided us the latest 
radio ephemeris data on the pulsar.
We are grateful to the AAO staffs in the recoating work of the 3.8m mirror.
The receipt of JSPS Research Fellowships (JH, AK, TN, GPR, KS, GJT and 
TY) is also acknowledged.
Finally, we thank the anonymous referee whose comments helped us improve
the manuscript.

\clearpage
\begin{deluxetable}{ccccccc}
\footnotesize
\tablecaption{
Summary of the observations and the analysis results.
The number of events in the ` After noise reduction' column are those
remaining after the 
$N_{hit}$ and $N_{p.e.}$ cuts are applied to obtain the
number of air shower events.
Flux upper limits for the 1996 and 1998 data are calculated as a
99.5$\%$ confidence level.
For the 1997 data, the results in each newmoon season are also presented
with the excess counts per minute. 
\label{obssum}}
\tablewidth{0pt}
\tablehead{ 
   &   &  Threshold 
 & \multicolumn{3}{c}{Number of Events} 
 & Flux or Upper Limit \\ 
 Observation   & Time &  Energy   &   &  After noise     &  After image  
 & ($\times 10^{-12} cm^{-2} s^{-1}$) \\
 Period   & (min)    &  (TeV)    &   Recorded  &  reduction &  selection &
} 
\startdata
1996 ON              & 1590  & 4.5 &  91622 &  16111 &  170 & $<$1.9        \nl
$~~~~~~~~$OFF        & 1590  &     &  99948 &  17297 &  169 &               \nl
1997 ON              & 1928  & 1.9 & 367689 & 106624 & 1388 & 2.9           \nl
$~~~~~~~~$OFF        & 1928  &     & 282156 & 106772 & 1180 &               \nl
1998 ON              & 1274  & 2.5 &  89752 &  26543 &  345 & $<$2.0        \nl
$~~~~~~~~$OFF        & 1274  &     &  90002 &  26705 &  309 &               \nl
                     &       &     &        &        &      & (excess/min)  \nl
March 1997 ON        &  345  &     &  73742 &  19440 &  261 & 0.10$\pm$0.06 \nl
$~~~~~~~~~~~~~~~$OFF &  345  &     &  62193 &  19610 &  227 &               \nl
April 1997 ON        &  598  &     & 101909 &  33504 &  426 & 0.12$\pm$0.05 \nl
$~~~~~~~~~~~~~~~$OFF &  598  &     &  82334 &  33610 &  381 &               \nl
May   1997 ON        &  985  &     & 192038 &  53680 &  701 & 0.13$\pm$0.04 \nl
$~~~~~~~~~~~~~~~$OFF &  985  &     & 137629 &  53552 &  572 &               \nl
 
\enddata
\end{deluxetable}
\clearpage

\begin{deluxetable}{ll}
\footnotesize
\tablecaption{Pulsar timing data (from radio observation) used in the 
periodicity analysis (\cite{man98}).
\label{tbl-eph}}
\tablewidth{0pt}
\tablehead{ 
 Parameter & Value
} 
\startdata
Validity range (MJD)       &  50114 -- 51094               \nl
$\nu_{0}$ (s$^{-1}$)       &  6.6244525661182              \nl
$\dot{\nu}_{0}$ (s$^{-2}$) &  -6.73155 $\times$ 10$^{-11}$ \nl
$\dot{\nu}_{0}$ (s$^{-3}$) &  1.95 $\times$ 10$^{-21}$     \nl
t$_{0}^{geo}$ (MJD)        &  50604.000000816              \nl
\enddata
\end{deluxetable}
\clearpage

\begin{deluxetable}{cccc}
%\begin{center}
\footnotesize
\tablecaption{Results of the periodicity analysis. 
The H-test statistics for each year are shown.
Because of the GPS clock problem (see text), the 1997 data are divided into
three observation seasons.
Chance probabilities P($>$H) are calculated against a uniform light curve 
(no periodicity).
The corresponding 3$\sigma$ flux upper limits are also shown. \label{tbl-1}}
\tablewidth{0pt}
\tablehead{ 
 Observation & \multicolumn{2}{c}{H-test} & flux upper limit\\ 
 Period      & H  & P($>$H) & ($\times 10^{-12} cm^{-2} s^{-1}$) 
} 
\startdata
1996       & 3.55 & 0.24 & 1.7 \nl
March 1997 & 6.37 & 0.08 & 5.2 \nl
April 1997 & 0.61 & 0.78 & 2.6 \nl
May   1997 & 0.84 & 0.71 & 2.1 \nl
1998       & 3.85 & 0.21 & 1.5 \nl
 
\enddata
%\end{center}
\end{deluxetable}

\clearpage

\figcaption[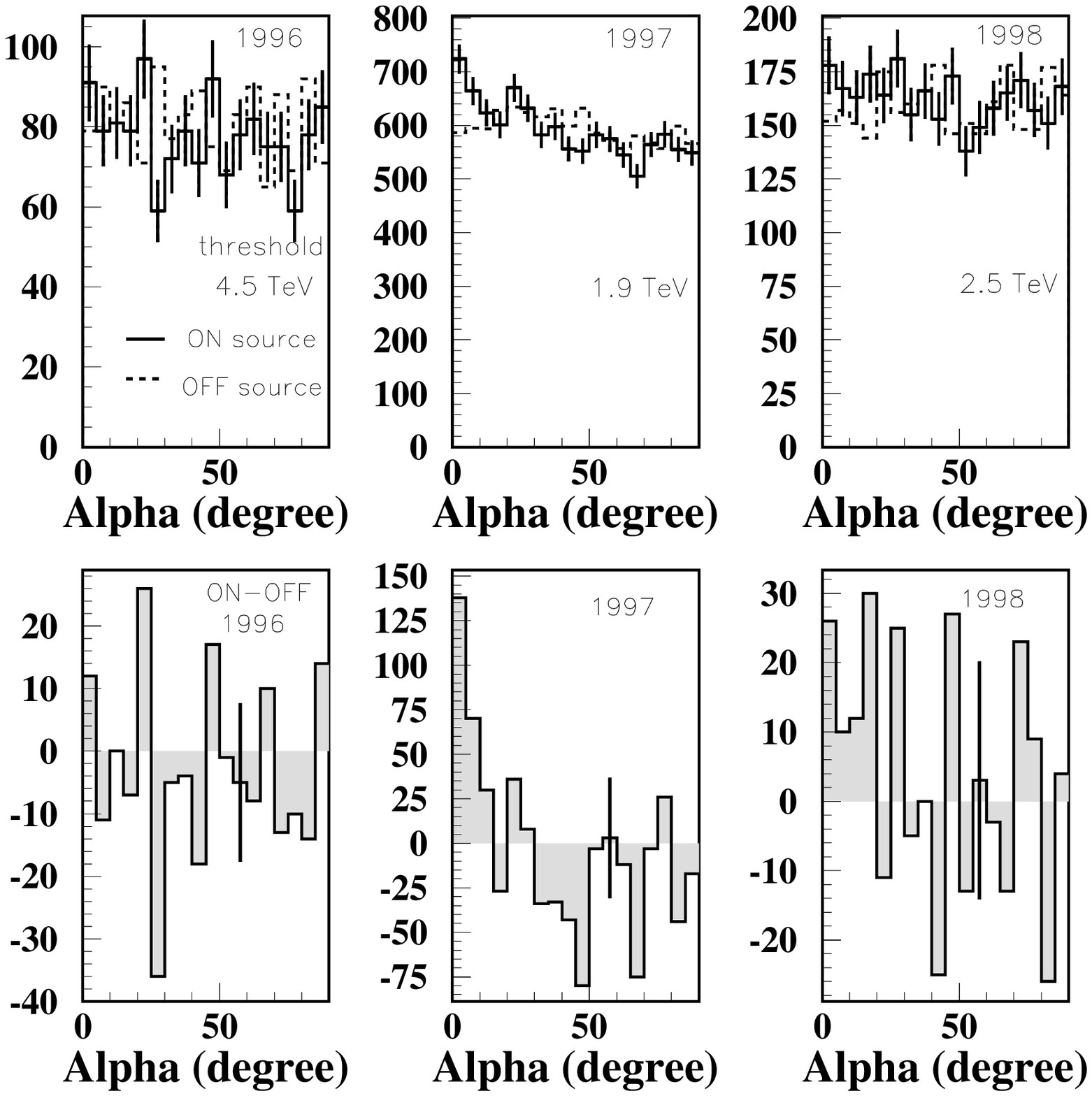]{
Distributions of the $\alpha$ parameter after all other image cuts. 
The solid and dashed lines in upper figures show the ON source and OFF source
results, respectively.
The bottom figures represent the ON--OFF counts of the upper figures.
\label{Fig1}}

\figcaption[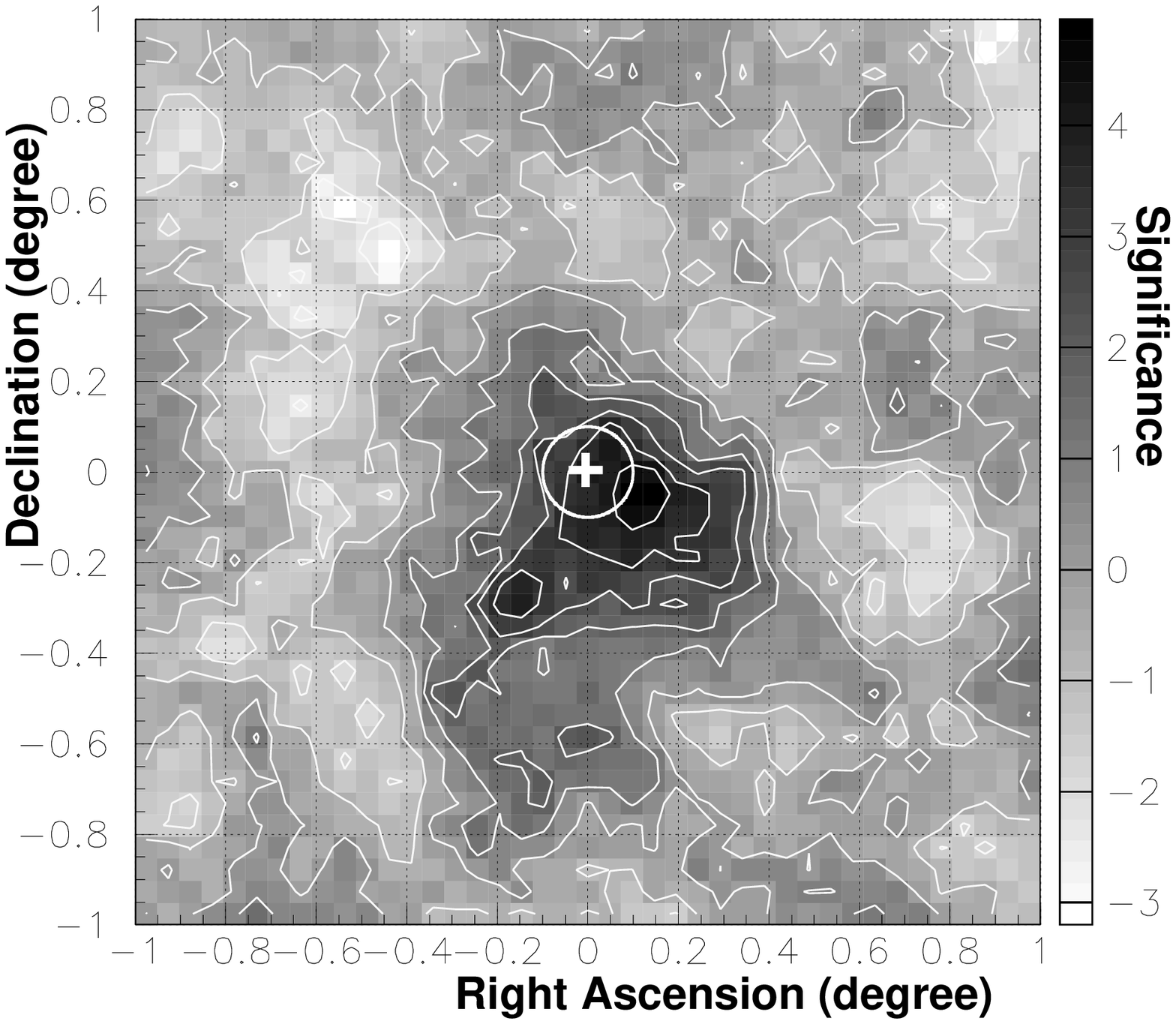]{
The contour map of the significance around the pulsar position in the 1997
data.
North is to the top of the figure, and west is to the right.
The field of view is $2^{\circ}\times2^{\circ}$ and the pulsar position is 
indicated by the cross.
The distance from the pulsar position to the peak of the excess (SW from the 
pulsar) is $0^{\circ}.1$ and is consistent with the pulsar position within
the source localization error, which is indicated by the circle.
\label{Fig4}}

\clearpage

\plotone{alpha5.eps}

\clearpage

\plotone{map97.eps}

\end{document}